# Meshfree simulation of multiphase flows with SPH family methods

Feiguo Chen[a,b,*]

[a]*State Key Laboratory of Multiphase Complex Systems, Institute of Process Engineering, Chinese Academy of Sciences, Beijing 100190, China*

[b]*Innovation Academy for Green Manufacture, Chinese Academy of Sciences, Beijing 100190, China*

[*]*Email address*: fgchen@ipe.ac.cn (F. Chen)

A B S T R A C T

With meshfree and fully Lagrangian features of particle methods, smoothed particle hydrodynamics (SPH) is suitable to achieve high-accurate simulations of multiphase flows with large interfacial deformations, discontinuities, and multi- physics. In this review, the basic concept of SPH is first briefly introduced. Then, various implementations of SPH in regard to multiphase flow simulations are summarized and discussed. Some problems associated with SPH simulations of multiphase flows are suggested as requiring attention.

## 1. Introduction

Multiphase flows exist widely in nature and in industrial applications. It is important to model and simulate the multiphase flows with high precision and efficiency. However, the complexities in the flow patterns and interface tracking limit the capability of traditional grid-based methods in multiphase flow simulations, especially with the occurrence of large interfacial deformations, discontinuities, and multi-physical aspects. And meshfree methods can deal well with the problems encountered in grid-based methods. Among the meshfree methods developed in the past decades, smoothed particle hydrodynamics (SPH) is well-known and widely applied.

SPH was originally invented in 1977 to simulate non-axisymmetric phenomena in astrophysics (Gingold and Monaghan 1977; Lucy 1977). It has been developed extensively and become one of the standard techniques to model fluids (Monaghan 2005). A considerable amount of literatures, including comprehensive reviews of SPH (Monaghan 1992; Li and Liu 2002; Liu and Liu 2003; Koumoutsakos 2005; Monaghan 2005; Cleary et al. 2007; Monaghan 2011; Shadloo et al. 2016; Violeau and Rogers 2016; Gotoh and Khayyer 2018; Ye et al. 2019), exist on the SPH method applied to a number of research topics, such as interfacial flow, rheology, instability, multiphase flow, breakage, elasticity and fracture, explosion, wave propagation, and heat conduction.

From a purely Lagrangian perspective, the SPH method offers the following main advantages (Li and Liu 2002; Violeau and Rogers 2016):

1) Ease of handling large deformations and rapidly moving free surfaces;

2) Capturing and tracing of discontinuities spontaneously, a feature which is difficult using grid-based methods;

3) Efficient application in simulations of multiphase flows exhibiting multi-physics features;

4) Well controlled numerical accuracy as it is easy to refine or coarsen the SPH particles needed;

5) Easy programming implementation (for both 2D and 3D simulations) and good scalability in parallel computing.

All the above features ensure the SPH method is particularly suitable for multiphase flow simulations with high accuracy, which is difficult or even unachievable using grid-based methods. Although the literatures cover many different multiphase flows treated in SPH simulations, and also many reviews have critiqued the various applications of the SPH methods, few have summarized the different methods associated with SPH implementation to multiphase flows. This paper mainly addresses and compares the various implementations.

This survey is organized as follows: Section 1 reviews the development, characteristics, and applications of SPH. In Section 2, a brief introduction of the SPH method is presented. Section 3 summarizes the various applications of SPH for multiphase flow simulations. A discussion of miscellaneous topics concerning the SPH method and multiphase flow simulations is presented in Section 4. Finally, a summary of the survey is provided in Section 5.

## 2. Methods of SPH

SPH is a typical Lagrangian particle method using a collection of particles with specified physical properties (e.g. density, viscosity, mass, velocity, and temperature) to represent the fluid (Monaghan 1992; Liu and Liu 2003; Monaghan 2005; Wang et al. 2016). The motion and changes in properties of the particles are governed by the conservation equations of mass, momentum, and energy. Their numerical approximation is obtained by weight integration of the kernel and particle interpolation.

### *2.1. Basic principles of SPH*

In brief, the basic idea of SPH is summarized as follows (Liu and Liu 2003). A collection of particles with physical properties are used to discretize the problem domain and change the governing equations of the fluid to ordinary differential equations with various boundary conditions. Then, a weight integral representation with a certain kernel approximation is employed to obtain numerical solutions. The key to the SPH method is the interpolating approximation, including the kernel approximation and the particle approximation.

2.1.1. Kernel approximation

The value of a continuous smooth function at a point $\boldsymbol{x}$ in the integral domain Ω is defined as (Liu and Liu 2003)

$$f(x) = \int_\Omega f(x')\delta(x - x')dx' , \tag{1}$$

where $\delta(x\text{-}x')$ is the Dirac delta function

$$\delta(x - x') = \begin{cases} \infty & x = x' \\ 0 & x = x' \end{cases} \tag{2}$$

To eliminate the appearance of infinite values arising from the delta function, a smoothing kernel function is introduced to approximate the delta function. Then, the function value is expressed as (Monaghan and Kocharyan 1995; Liu and Liu 2003)

$$\langle f(x) \rangle = \int_\Omega f(x')W(x - x', h)dx' , \tag{3}$$

where $W(x\text{-}x', h)$ is the kernel function with $h$ standing for the smoothing length metering the influencing area of the weight integration.

The function derivation is approximately performed with the kernel function as

$$\langle \nabla \cdot f(x) \rangle = \int_\Omega \left[\nabla \cdot f(x')\right] W(x - x', h)dx' . \tag{4}$$

According to Gauss integral formula, the above expression can be rewritten as

$$\langle \nabla \cdot f(x) \rangle = -\int_\Omega f(x') \cdot \nabla W(x - x', h)dx' . \tag{5}$$

2.1.2. Particle approximation

The domain is discretized into a collection of particles (Fig. 1), and the volume of the infinitesimal element at particle $j$ equals the volume of the particle, $\Delta V_j$. If $\rho_j$ is the density of particle $j$, its mass is

$$m_j = \rho_j \Delta V_j . \tag{6}$$

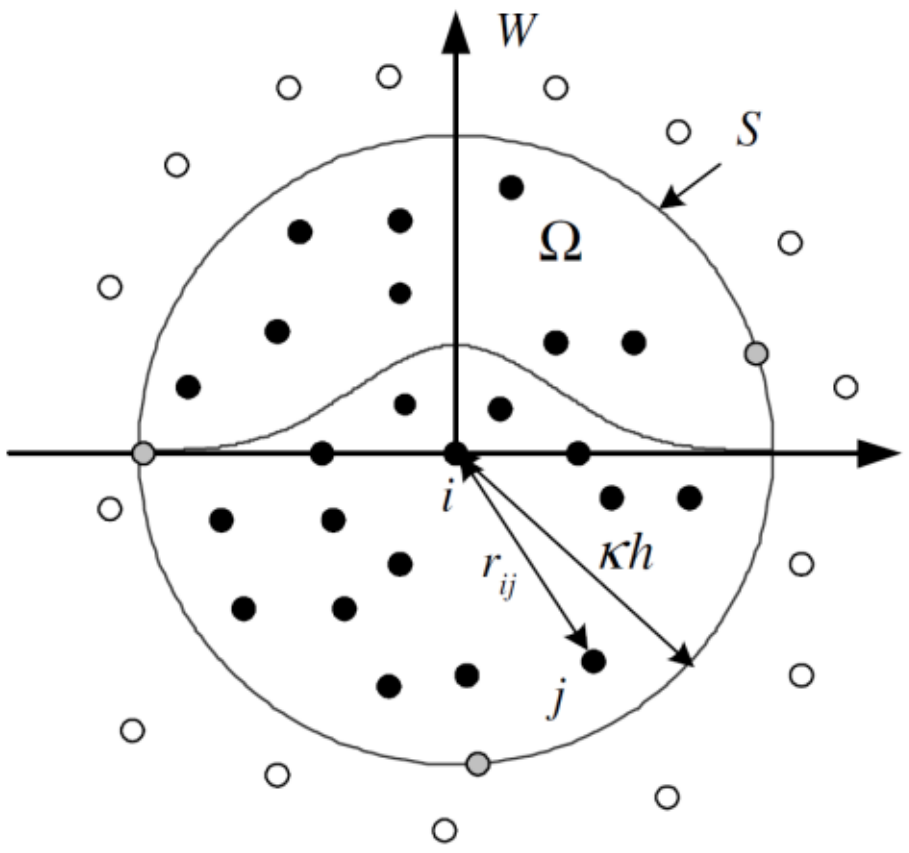

**Fig. 1.** Schematic illustration of SPH particle approximation (Liu and Liu 2003)

The integral formula of the SPH kernel approximation is expressed as (Liu and Liu 2003)

$$\langle f(x)\rangle = \sum_j \frac{m_j}{\rho_j} f(x_j) W(x-x_j,h) . \tag{7}$$

The value of the function at particle *i* is written

$$\langle f(x_i)\rangle = \sum_j \frac{m_j}{\rho_j} f(x_j) W_{ij} , \tag{8}$$

where $W_{ij}=W(x_i\text{-}x_j,\ h)$. The function derivation is expressed within the particle approximation as

$$\langle \nabla\cdot f(x)\rangle = -\sum_j \frac{m_j}{\rho_j} f(x_j)\cdot \nabla W(x-x_j,h) , \tag{9}$$

and the derivation at particle *i* is written

$$\langle \nabla\cdot f(x_i)\rangle = -\sum_j \frac{m_j}{\rho_j} f(x_j)\cdot \nabla W_{ij} = \sum_j \frac{m_j}{\rho_j} f(x_j)\cdot \nabla_i W_{ij} . \tag{10}$$

The following two identities are commonly used to transform the gradient operator using the density (Monaghan 1992; Monaghan 2005; Liu and Liu 2010)

$$\nabla\cdot f(x) = \frac{1}{\rho}\left[\nabla\cdot(\rho f(x)) - f(x)\cdot\nabla\rho\right] \tag{11}$$

$$\nabla\cdot f(x) = \rho\left[\nabla\cdot\left(\frac{f(x)}{\rho}\right) + \frac{f(x)}{\rho^2}\cdot\nabla\rho\right] . \tag{12}$$

Hence, the results for the derivation at particle *i* are calculated as

$$\langle \nabla \cdot f(x_i) \rangle = \frac{1}{\rho_i}\left[ \sum_j m_j \left( f(x_j) - f(x_i) \right) \cdot \nabla_i W_{ij} \right] \tag{13}$$

$$\langle \nabla \cdot f(x_i) \rangle = \rho_i \left[ \sum_j m_j \left( \frac{f(x_i)}{\rho_i^2} + \frac{f(x_j)}{\rho_j^2} \right) \cdot \nabla_i W_{ij} \right]. \tag{14}$$

That is, the value of the function and its derivation can be obtained by a weighted average of discrete particles.

*2.2. Discretization of governing equations*

The fluid dynamics are governed by the conservation equations, namely, the continuity equation, the momentum equation (Navier–Stokes equation), and the energy equation, which are presented in the form

$$\begin{aligned} \frac{d\rho}{dt} &= -\rho \nabla \cdot \boldsymbol{u} \\ \frac{d\boldsymbol{u}}{dt} &= \frac{1}{\rho}\nabla \cdot \boldsymbol{S} + \boldsymbol{g} \\ \frac{de}{dt} &= \frac{1}{\rho}\boldsymbol{S} : \nabla \boldsymbol{u} - \frac{1}{\rho}\nabla \cdot \boldsymbol{q} \end{aligned} \tag{15}$$

where $\boldsymbol{u}$ is the velocity, $\rho$ the density, $\boldsymbol{S}$ the stress tensor, $\boldsymbol{g}$ the body force per unit mass, $e$ the specific internal energy, and $\boldsymbol{q}$ the heat flux vector. The stress tensor $\boldsymbol{S}$ takes the explicit form

$$\boldsymbol{S} = -p\boldsymbol{I} + \boldsymbol{\sigma}, \tag{16}$$

where $\boldsymbol{I}$ denotes the unit tensor, $p$ the internal pressure, and $\boldsymbol{\sigma}$ the viscous stress expressed as

$$\boldsymbol{\sigma} = \mu(\nabla \boldsymbol{u} + \boldsymbol{u}\nabla) + \left( \xi - \frac{2}{3}\mu \right)(\nabla \cdot \boldsymbol{u})\boldsymbol{I}, \tag{17}$$

where $\mu$ and $\xi$ are the shear and bulk viscosity coefficients respectively.

The SPH representations for the conservation equations are written (Monaghan 1992; Liu and Liu 2003; Liu and Liu 2010; Yang et al. 2014)

$$
\begin{aligned}
\frac{d\rho_i}{dt} &= \sum_j m_j \left(\boldsymbol{u}_j - \boldsymbol{u}_i\right)\cdot\nabla_i W_{ij} \\
\frac{d\boldsymbol{u}_i}{dt} &= \sum_j m_j \left(\frac{\boldsymbol{S}_i}{\rho_i^2} + \frac{\boldsymbol{S}_j}{\rho_j^2}\right)\cdot\nabla_i W_{ij} + \boldsymbol{g}_i \\
\frac{de_i}{dt} &= \frac{1}{2}\sum_j m_j \left(\frac{\boldsymbol{S}_i}{\rho_i^2} + \frac{\boldsymbol{S}_j}{\rho_j^2}\right):\left(\boldsymbol{u}_j - \boldsymbol{u}_i\right)\nabla_i W_{ij} \\
&\quad - \sum_j m_j \left(\frac{\boldsymbol{q}_i}{\rho_i^2} + \frac{\boldsymbol{q}_j}{\rho_j^2}\right)\cdot\nabla_i W_{ij}
\end{aligned}
\tag{18}
$$

In the SPH method, there are two forms to calculate the particle density. One uses the derivation as

$$
\frac{d\rho_i}{dt} = \sum_j m_j \left(\boldsymbol{u}_j - \boldsymbol{u}_i\right)\cdot\nabla_i W_{ij} \ . \tag{19}
$$

The other represents the particle density as a weight sum of other particles,

$$
\rho_i = \sum_j m_j W_{ij} \ . \tag{20}
$$

It should be noted that the continuity and momentum equations are not closed with respect to the unknown pressure expression in the momentum equation. The pressure of the fluid is obtained from the equation of state (EOS) generally. For the incompressible SPH, the Poisson equation for pressure is solved using the projection method to couple the velocity and pressure.

*2.3. EOS*

In the SPH method, the EOS of the fluid is an important factor in calculating the pressure term in the momentum equation effectively. The early SPH simulation (Monaghan 1994) used the EOS proposed by Batchelor (1967)

$$
p = p_0\left[\left(\frac{\rho}{\rho_0}\right)^{\gamma} - 1\right], \tag{21}
$$

where $\gamma$ is a constant ranging from 1 to 7; the subscript zero signifies reference quantities. It is also the most commonly-used EOS for weakly compressible SPH. For large $\gamma$, the high sensitivity of pressure on the density may result in an extremely large error in pressure with a large density fluctuation.

Another commonly-used EOS takes the simple form (Morris et al. 1997; Liu and Liu 2005)

$$
p = c^2\rho, \tag{22}
$$

where $c$ is the speed of sound for the fluid; it is a key factor in balancing the approximation for real fluids and the efficiency of the time evolution.

For a van der Waals (vdW) fluid, the EOS used is (Silbey et al. 2004)

$$p = \frac{nk_B T}{1 - nb} - an^2 \tag{23}$$

or

$$p = \frac{\rho \bar{k} T}{1 - \rho \bar{b}} - \bar{a}\rho^2, \tag{24}$$

where $n$ is the number density, $T$ the temperature, $k_B$ the Boltzmann's constant, $a$ and $b$ are coefficients denoting the cohesive interaction and molecular size, respectively, and $\bar{k} = k_B/m$, $\bar{a} = a/m^2$, $\bar{b} = b/m$ with $m$ for particle mass. The second term is the cohesive interaction for forming a stable surface for droplets. The vdW EOS is also called the cohesive pressure model.

*2.4. Kernel function*

One of the central issues for the SPH method is how to perform an accurate and stable approximation effectively. The smoothing kernel function, which is used for the kernel and particle approximation, determines the interpolation pattern and the weighted integration.

Many researchers have investigated the smoothing kernel and proposed a variety of kernel functions. For all smoothing kernel functions, the general requirements that can be applied are (Liu and Liu 2003; Monaghan 2005; Liu and Liu 2010; Wang et al. 2016):

(1) It must be normalizable over its domain of support (Unity),

$$\int_\Omega W(x - x', h)\, dx' = 1; \tag{25}$$

(2) It must be compactly supported (Compactness),

$$W(x - x', h) = 0, \quad \text{for } |x - x'| > \kappa h; \tag{26}$$

(3) Its value within the support domain must be positive,

$$W(x - x', h) \geq 0; \tag{27}$$

(4) It should satisfy the Dirac delta function condition as the smoothing length approaches zero (Delta function property),

$$\lim_{h \to 0} W(x - x', h) = \delta(x - x'); \tag{28}$$

(5) It should be decreasing, even and sufficiently smooth..

In theory, any function satisfying the above requirements can be used as the smoothing kernel function for the SPH method. Its general construction was discussed in (Liu and Liu 2010). A commonly used kernel function is the Gaussian function originally presented in (Gingold and Monaghan 1977)

$$W(s,h) = \alpha_d e^{-s^2}, \tag{29}$$

and the most widely used one is the piecewise spline function, such as the cubic spline function (Monaghan and Pongracic 1985; Liu and Liu 2005).

$$W(s,h) = \alpha_d \begin{cases} s^3 - 6s + 6 & 0 \le s < 1 \\ (2-s)^3/6 & 1 \le s < 2 \\ 0 & s \ge 2 \end{cases}, \tag{30}$$

where $s=|\mathbf{x}-\mathbf{x}'|/h$, $\alpha_d$ is the normalization factor related to the dimension $d$. More details on typical smoothing kernel functions can be found in review literature (Liu and Liu 2003; Liu and Liu 2010; Wang et al. 2016).

*2.5. Surface tension*

The surface tension, occurring at the interface of two phases, is an important factor for multiphase flows, in particular, gas–liquid flows, micro-flows, and surface flows, where it is the dominant factor influencing flow behaviors. An accurate numerical expression of surface tension is critical in modeling multiphase flows well.

From the viewpoint of molecular interaction, surface tension derives from asymmetry in the inter-molecular forces at the interface (Silbey et al. 2004); see Fig. 2.

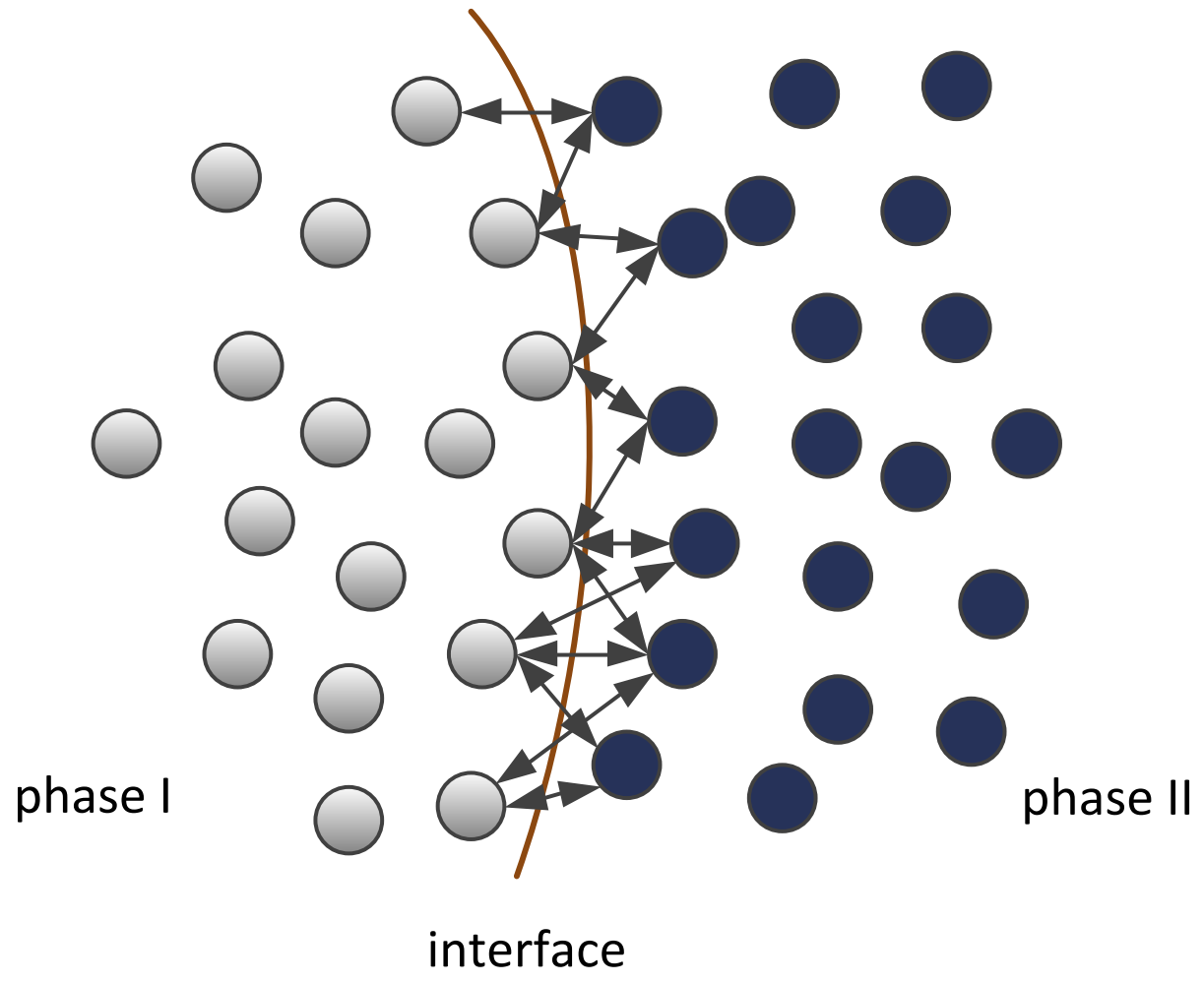

**Fig. 2.** Schematic illustration of the interaction of molecules at the interface

In SPH methods, there are three typical models that describe surface tension, namely, the continuum surface force (CSF) model, the inter-particle force (IPF) model, and the cohesive pressure model (CPM). Sometimes, CPM is regarded as a generalized IPF.

2.5.1. CSF model

The CSF model is a macroscopic force model that describes surface tension as a biased molecular interaction between two phases. It was first proposed in (Brackbill et al. 1992). The interface between the fluids is viewed as a transition region of finite thickness. The force density of the surface tension is defined proportional to the local curvature of the surface (Morris 2000; Nomura et al. 2001; Zhang 2010; Lind et al. 2012) and is expressed as

$$\boldsymbol{f}_s = \gamma\kappa\boldsymbol{n} + \nabla_s\sigma, \tag{31}$$

where $\gamma$ is the surface tension coefficient, $\kappa$ the local curvature of the interface, $\boldsymbol{n}$ the unit normal to the interface, and $\nabla_s\sigma$ the tangential term from the difference of the surface tension at the interface.

Generally, a color function (Brackbill et al. 1992; Morris 2000) is used to distinguish different phases, defined as

$$C_i = \begin{cases} s & \text{particle } i \in \text{phase } s \\ 0 & \text{else} \end{cases}. \tag{32}$$

The interface is then defined as a finite transitional band with non-zero color gradient. The normal vector of the interface is calculated from

$$\boldsymbol{n} = \frac{\nabla C}{|\nabla C|}. \tag{33}$$

The surface tension force is finally converted into a body force,

$$\boldsymbol{f} = -\gamma(\nabla\cdot\boldsymbol{n})\nabla C. \tag{34}$$

2.5.2. IPF model

The IPF model uses an implicit function representing the surface tension with pair-wise symmetric forces on the interfacial SPH particles. Generally, the IPF is expressed in the form

$$f_{ij} = f(\boldsymbol{r}_i - \boldsymbol{r}_j, h), \tag{35}$$

when particles $i$ and $j$ belong to two different phases in the transition region. The additional force on the particles interacting with the other phases results in a net acceleration produced by surface tension,

$$\left(\frac{d\boldsymbol{v}_i}{dt}\right)_s = \frac{1}{m_i}\sum_j \boldsymbol{f}_{ij}. \tag{36}$$

Commonly used IPF models in the literature include the Lennard-Jones force model (Zhang et al. 2008)

$$\boldsymbol{f}_{ij} = -\frac{m_j}{\rho_i} 4\varepsilon \left[ \left( \frac{\sigma_r}{r_{ij}} \right)^{12} - \left( \frac{\sigma_r}{r_{ij}} \right)^{6} \right] \left( \nabla W_{ij} \right) \frac{\boldsymbol{r}_{ij}}{r_{ij}}, \tag{37}$$

the cosine force model (Tartakovsky and Meakin 2005)

$$\boldsymbol{f}_{ij} = -s \cos\left( \frac{3\pi}{2\kappa h} r_{ij} \right) \frac{\boldsymbol{r}_{ij}}{r_{ij}}, \tag{38}$$

the cubic polynomial force model (Su et al. 2013)

$$\boldsymbol{f}_{ij} = s \left[ 1 - \frac{57}{48} \left( \frac{r_{ij}}{h} \right)^2 + \frac{45}{96} \left( \frac{r_{ij}}{h} \right)^3 \right] \frac{\boldsymbol{r}_{ij}}{r_{ij}}, \tag{39}$$

and the simple repulsive force model (Zhou et al. 2008)

$$\boldsymbol{f}_{ij} = s \frac{\boldsymbol{r}_{ij}}{r_{ij}^2}. \tag{40}$$

More IPF models are given in the summary by (Wang et al. 2016).

2.5.3. CPM

The CPM can be considered as a generalized IPF, in which the additional pair-wise force on the interfacial particles is implicitly projected by the cohesive EOS. CPM is based on the vdW EOS [Eq. (24)].

According to (Nugent and Posch 2000), the interaction range of cohesive pressure is assumed to exceed that of other forces. Thus, the cohesive pressure term for internal particles is balanced to lead to a zero net force; and the unbalanced cohesive pressure in the interfacial region results in a surface tension arising from the attractive term ($\bar{a}\rho^2$) in the vdW EOS. The net acceleration of particles within the interfacial region has the form

$$\frac{d\boldsymbol{v}_i}{dt} = 2\bar{a} \sum_j m_j \nabla_i W_{ij}^H, \tag{41}$$

where $W_{ij}^H = W(\boldsymbol{r}_i - \boldsymbol{r}_j, H)$ is the kernel function with smoothing length $H \geqslant 2h$.

*2.6. Solid boundary conditions*

As in other discrete methods, the implementation of the solid boundary conditions in the SPH method is not as straightforward as in the grid-based methods. The objectives of the treatment of the boundary conditions in SPH are to prevent fluid particles penetrating the solid walls, and to exert a weighted interpolation at the boundary while the smoothing approximation is truncated, and then to apply the free-slip, non-slip or partial slip boundary conditions as required. In principal, boundary treatment techniques in discrete methods (including SPH method) can be roughly classified into two categories, namely, particle representation and geometrical representation.

2.6.1. Particle representation

Within the particle representation, the solid boundary is represented by a collection of fixed particles, and an additional force between fluid particles and boundary particles is applied to produce the specified boundary conditions. The boundary particle force model (Monaghan 1994), the image particle model (Morris et al. 1997), and the dynamic particle model (Liu and Liu 2003; Gong et al. 2009) are three typical models distinguished by the way the solid particles are generated.

In the boundary particle force model, solid virtual particles are positioned right on the boundary (Fig. 3). Repulsive forces from the solid particles acting on the fluid particles are employed to prevent non-physical penetration of the fluid particles moving toward the boundary. Researchers have proposed various repulsive force models to improve and correct the boundary force computation (Monaghan 1994; Shadloo et al. 2016; Wang et al. 2016).

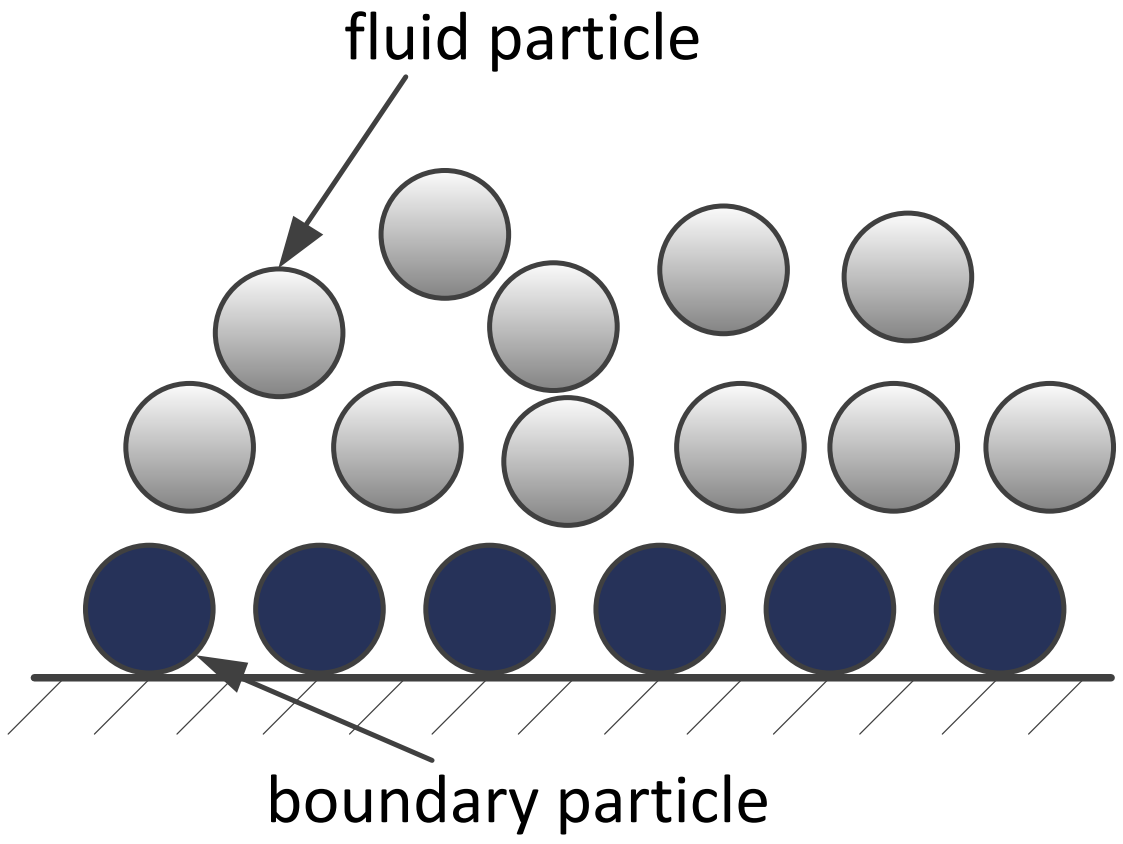

**Fig. 3.** Schematic illustration of solid virtual particles

The image particle model is also called the ghost or mirror particle model (Fig. 4), in which the image particles are generated by mirroring fluid particles along the solid boundary (Morris et al. 1997; Bierbrauer et al. 2009). The locations and velocities of the image particles are in tangential symmetry to the fluid particles; their other properties are retained. Therefore, image particles are dynamically re-generated every timestep.

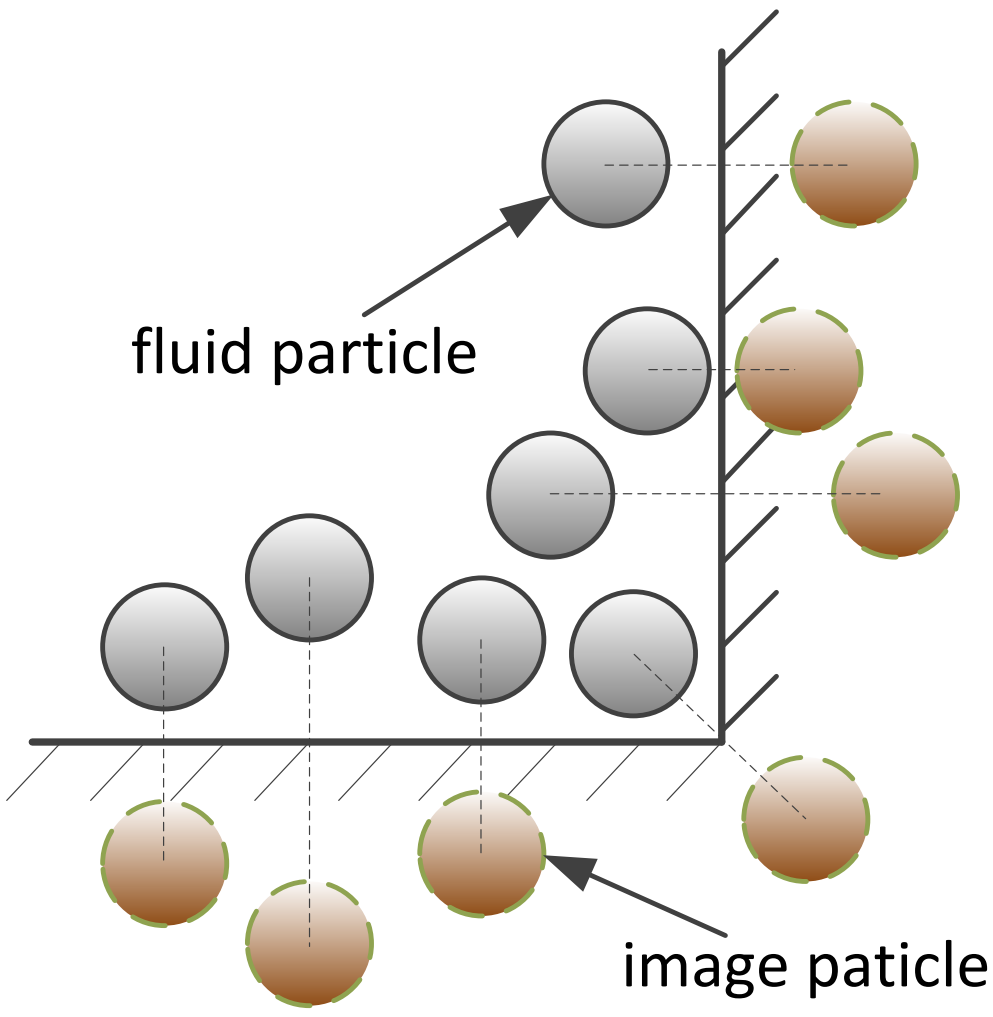

**Fig. 4.** Schematic illustration of image particles

In the dynamic particle model (Fig. 5), the solid boundary is represented by a collection of fixed particles along the boundary (Liu and Liu 2003; Gómez-Gesteira and Dalrymple Robert 2004; Gong et al. 2009). The locations of boundary particles are constant with time but other properties are obtained by linear extrapolation from fluid particles close to the boundary. These boundary particles provide the kinetic boundary conditions.

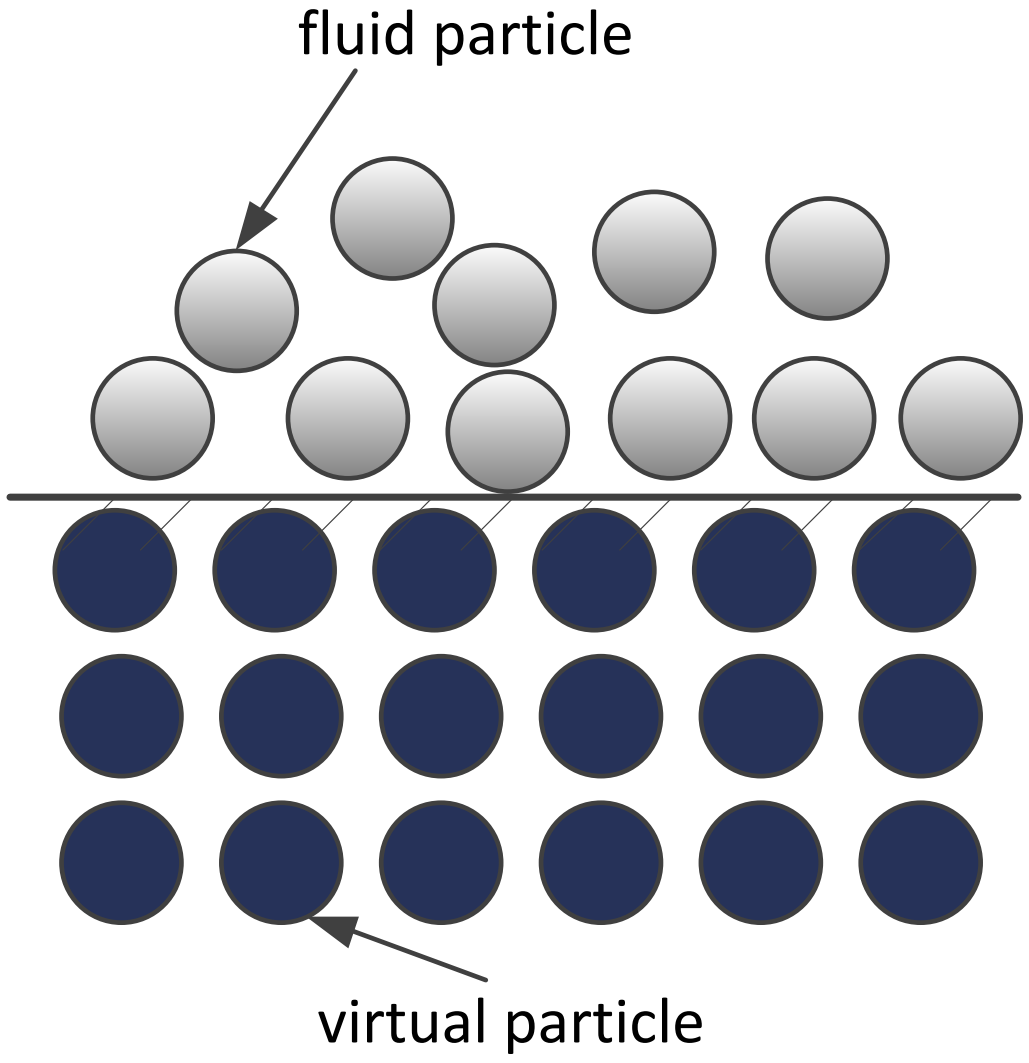

**Fig. 5.** Schematic illustration of dynamics particles along the boundary

2.6.2. Geometrical representation

Within the geometrical representation, the semi-analytical model (Kulasegaram et al. 2004; Ferrand et al. 2013; Mayrhofer et al. 2015) is introduced. In the semi-analytical model (Fig. 6), the boundary properties (e.g., segment and inward normal) are redefined by the geometrical representative particles (Ferrand et al. 2013).

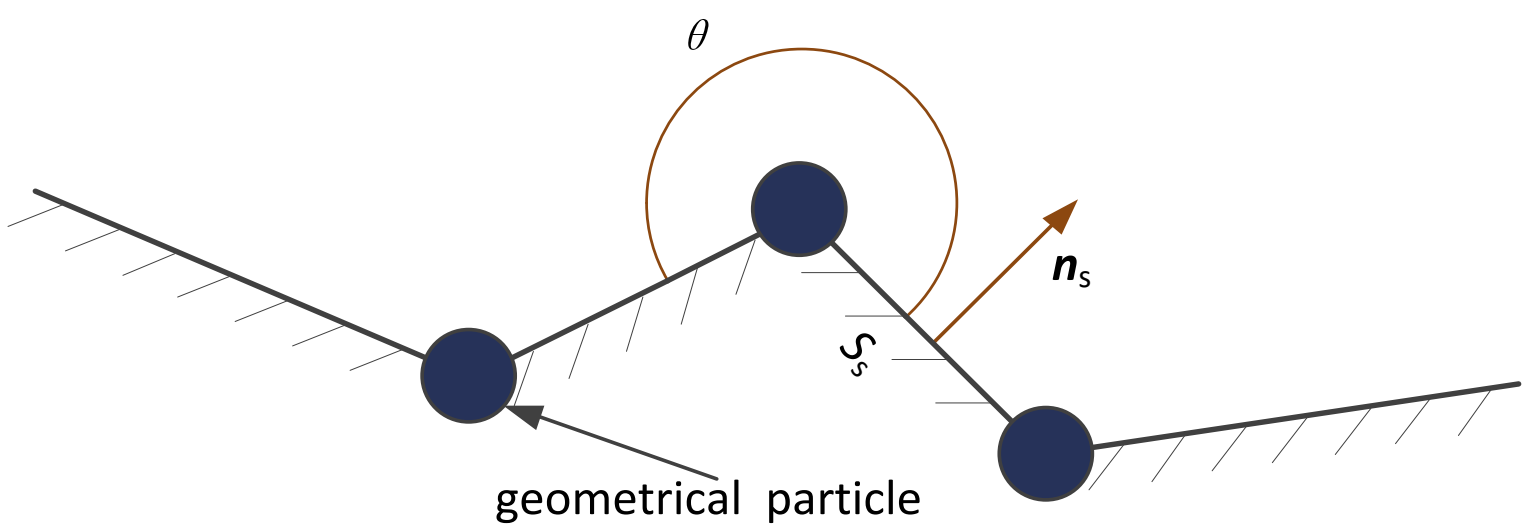

**Fig. 6.** Schematic of boundary property definitions (Ferrand et al. 2013).

A renormalizing factor is introduced to reconstruct the governing equations for the fluid particle close to the boundary. The operator of the gradient is represented in the form (Ferrand et al. 2013)

$$\nabla f_i = \frac{\rho_i}{\gamma_i}\sum_{j\in F} m_j\left(\frac{f_i}{\rho_i^2}+\frac{f_j}{\rho_j^2}\right)\nabla W_{ij} - \frac{\rho_i}{\gamma_i}\sum_{s\in S} m_j\left(\frac{f_i}{\rho_i^2}+\frac{f_s}{\rho_s^2}\right)\rho_s \nabla\gamma_{is}\,. \tag{42}$$

The correction parameters in the above operator are defined as

$$\gamma_i = \int_\Omega W\left(\boldsymbol{r}-\boldsymbol{r}_i\right)dr \tag{43}$$

$$\nabla\gamma_{as} = \left(\int_{r_{e1}}^{r_{e2}} W\left(\boldsymbol{r}-\boldsymbol{r}_i\right)dl\right)\boldsymbol{n}\,. \tag{44}$$

### *2.7. Some variants of the SPH method for multiphase flow*

#### 2.7.1. ISPH

In the conventional SPH method, the EOS, which describes the relationship between pressure and density, is used to represent the pressure term in the momentum equation. Hence, the fluid is considered compressible without the divergence-free velocity constraint. In (Cummins and Rudman 1999), the projection method is applied to the Poisson equation of pressure to address the divergence-free velocity in the SPH method and is implemented to establish the incompressible SPH (ISPH) method.

An intermediate velocity field $\boldsymbol{u}^*$ is calculated by integrating the momentum equation forward in time without the pressure gradient term. The pressure Poisson equation is then developed for the intermediate velocity field and solved to obtain the pressure required to enforce incompressibility,

$$\nabla\cdot\left(\frac{1}{\rho}\nabla p\right)_i = \frac{\nabla\cdot\boldsymbol{u}_i^*}{\Delta t}\,. \tag{45}$$

Thus, the velocity field is corrected using the pressure to achieve a divergent-free condition,

$$\boldsymbol{u}_i^{t+\Delta t} = \boldsymbol{u}_i^* - \Delta t \sum_j m_j \left( \frac{p_i}{\rho_i^2} + \frac{p_j}{\rho_j^2} \right) \nabla_i W_{ij} \ . \tag{46}$$

In (Ellero et al. 2007), a kinematic constraint imposed on the constant volume of fluid particles was introduced to achieve incompressibility. In (Hu and Adams 2007), the projection method for multiphase flows was implemented to establish the multiphase ISPH. In (Szewc et al. 2012), the implementations of the grid-based Poisson solver and the particle-based Poisson solver in the ISPH method were compared. Cornelis et al. (2014) presented an implicit ISPH to provide a realistic rendering of water flow. Aly (2015) applied the ISPH method to simulate the natural convection flow in a square and cubic cavity. Lind and collaborators used the ISPH method to simulate the free surface flows (Lind et al. 2012) and droplet deformation in air (Lind et al. 2016).

2.7.2. MPS

Koshizuka (1995) proposed another meshfree method based on the Lagrangian particle–moving particle semi-implicit (MPS) method for hydrodynamics and nuclear mechanics. With further developments, MPS has been widely used to deal with various problems such as dam breaking (Koshizuka and Oka 1996), breaking wave (Koshizuka et al. 1998), gas–liquid two-phase flow (Yoon et al. 1999; Yoon et al. 2001), and free surface flow (Ataie-Ashtiani and Farhadi 2006). Khayyer and Gotoh (2009) and Kondo and Koshizuka (2011) presented modifications of the MPS representation of pressure to eliminate its high-frequent unphysical numerical oscillations.

The MPS method has features similar to other particle methods, and a formulation of terms appearing in the governing equations that are quite similar to the SPH method. Both the ISPH and MPS methods employ the similar semi-implicit numerical scheme to solve incompressible flow, the major difference between them is in the use of the kernel function (Shao and Gotoh 2005). In the MPS method, a basic short-range kernel function (Koshizuka 1995) is often employed,

$$W(r,h) = \begin{cases} \dfrac{h}{r} - 1 & 0 < \dfrac{r}{h} \le 1 \\ 0 & \dfrac{r}{h} > 1 \end{cases} . \tag{47}$$

Its value goes to infinity at $r$=0 to avoid clustering of particles, corresponding to a finite value for the SPH kernel function at $r$=0.

2.7.3. MaPPM

The macro-scale pseudo-particle method (MaPPM) is another variant of the SPH method, proposed in (Ge and Li 2001; Ge and Li 2003) to provide high-accuracy simulations for particle–fluid systems. Instead of the derivative calculation of the kernel function, the derivative at a point in the MaPPM method is directly calculated as a weighted average of the finite difference at the point and its neighboring points,

$$\nabla f\,|_i = dim\sum_j \frac{f_{ji}}{r_{ij}^2}\boldsymbol{r}_{ij}\frac{m_j}{\rho_j}W_{ij}\ . \tag{48}$$

This replacement brings two major benefits: 1) A concise weighted average for the derivative can be maintained when conducting an intuitive simulation and direct analysis of multiphase flows; and 2) in calculations having no derivatives, other candidates can be adopted as the kernel function including non-differential functions.

In (Zhou et al. 2008), the IPF model proposed for surface tension and the boundary particle model in the MaPPM, and investigated droplet deformation and wetting behaviors of fluids on a wall. Ma et al. (2006) describes high-resolution simulations of a gas–solid suspension using MaPPM with 1024 particles represented by a cluster of "frozen particles", and investigated dynamic multi-scale structures. Xiong and collaborators (2010; 2011) employed hundreds of thousands of solid particles to investigate fluid–particle fluidization.

## 3. Various implementations of SPH for multiphase flow simulation

From previous literature on SPH applied to multiphase flows, the various implementations can be approximately classified into four categories: 1) Lagrangian solver for the two-fluid model (TFM); 2) multiphase SPH; 3) coupling of SPH and other discrete methods; and 4) coupling of SPH and grid-based methods.

### *3.1. Lagrangian solver for TFM*

TFM (Stewart 1979; Ishii and Mishima 1984; Gidaspow 1994) was developed as a continuum description for two phase flow with a distinct interface or dispersed bodies (bubbles or solid particles). Each phase is described separately, each having a set of conservation equations of mass, momentum, and energy. Interactions terms are used in the equations to couple mass, momentum, and energy transport across the interphases.

For fluid–solid two-phase flows, the governing equations are presented in the form (Gidaspow 1994):

fluid phase:

$$\frac{\partial \theta_f}{\partial t} + \nabla\cdot\left(\theta_f \boldsymbol{u}_f\right) = 0\,, \tag{49}$$

$$\frac{\partial\left(\theta_f \boldsymbol{u}_f\right)}{\partial t} + \nabla\cdot\left(\theta_f \boldsymbol{u}_f \boldsymbol{u}_f\right) = -\frac{\nabla p_f}{\rho_f} + \frac{\beta}{\rho_f}\left(\boldsymbol{u}_s - \boldsymbol{u}_f\right) + \frac{1}{\rho_f}\left(\nabla\cdot\mu_f\theta_f\nabla\right)\boldsymbol{u}_f + \theta_f \boldsymbol{g} \tag{50}$$

solid phase:

$$\frac{\partial \theta_s}{\partial t} + \nabla \cdot (\theta_s \boldsymbol{u}_s) = 0, \tag{51}$$

$$\begin{aligned} &\frac{\partial(\theta_s \boldsymbol{v}_s)}{\partial t} + \nabla \cdot (\theta_s \boldsymbol{u}_s \boldsymbol{u}_s) = \\ -\frac{\nabla p_s}{\rho_s} + \frac{\beta}{\rho_s}(\boldsymbol{u}_g - \boldsymbol{u}_s) &+ \frac{1}{\rho_s}(\nabla \cdot \mu_s \theta_s \nabla)\boldsymbol{u}_s + \theta_s \boldsymbol{g} \end{aligned} \tag{52}$$

where $\theta$ is the volume fraction, $\beta$ the coefficient of drag, and subscripts $f$ and $s$ signify fluid phase and solid phase, respectively. The constraints between the two sets of equations are $\theta_f + \theta_s = 1$ and $0 \le \theta_f$, $\theta_s \le 1$. The interaction coupling the two phases is described in terms of a drag term in the momentum equations.

As for single-phase flow, the SPH method can also be applied as the Lagrangian solution of the TFM, for which two phases are discretized into two separated sets of SPH particles. The distribution of the SPH particles of two phases naturally satisfies the constraint conditions $\theta_f + \theta_s = 1$ and $0 \le \theta_f$, $\theta_s \le 1$; and similar to other methods, the key to solving two phase problems is the interphase interaction and its discrete representation.

In (Monaghan and Kocharyan 1995), the two-phase flow for dusty gas is formulated using SPH. The interphase interaction term in the momentum equation is discretized as the drag kernel,

$$\frac{1}{dim} \sum_j m_j \frac{\beta_{ij}}{\rho_j} \left( \frac{\boldsymbol{u}_{ij} \cdot \boldsymbol{r}_{ij}}{r_{ij}^2 + \eta^2} \right) \boldsymbol{r}_{ij} W_{ij} \ , \tag{53}$$

where $\eta^2$ is a clipping constant ~$0.001h^2$ to prevent singularities with $r_{ij}^2=0$. The coefficient of drag $\beta$ depends on the SPH particles between gas and dust phases and is expressed as

$$\beta = \frac{2\rho_g \theta_s C_D}{d_p} \left| \boldsymbol{u}_g - \boldsymbol{u}_s \right|, \tag{54}$$

where $d_p$ is the diameter of dust particles, and $C_D$ is the standard coefficient of drag which depends on the Reynolds number $Re_p = \rho_g \mu_g |\boldsymbol{u}_g - \boldsymbol{u}_s| / d_p$ of the particle. The linear and angular momenta along the lines of centers are conserved for the interaction. The key feature of this approach is in the direct formulation of the interaction (drag) term.

Subsequently, Monaghan (Monaghan 1997) developed an implicit drag model to improve the accuracy and stability of the SPH computation. The velocity of particle $i$ is updated using the deceleration due to drag,

$$\boldsymbol{u}_i = \boldsymbol{u}_i^0 - \frac{\Delta t \boldsymbol{r}_{ji} m_i s_{ij} \left( \boldsymbol{u}_{ij}^0 \cdot \boldsymbol{r}_{aj} \right)}{1 + \Delta t \left( m_i + m_j \right) s_{ji} r_{ij}^2}, \tag{55}$$

where superscript 0 signifies the initial value before applying the drag term, $\Delta t$ the time-step, and the coefficient $s_{ij}$ is expressed as

$$s_{ij} = \frac{1}{dim} \frac{\beta_{ij}}{\rho_i \rho_j} \left( \frac{W_{ij}}{r_{ij}^2 + \eta^2} \right). \tag{56}$$

Laibe and Price (Laibe and Price 2012) adopted the variable smoothing length in the drag kernel function to perform simulations of gas–dust mixtures. Xiong et al. (2011) introduced a relationship between solid viscosity and solid volume fraction,

$$\mu_s = 5.0 \theta_s, \tag{57}$$

to simulate liquid–solid suspensions.

### *3.2. Multiphase SPH*

The SPH method has shown great potential in flow simulations. Researchers performing multiphase flows have extended this method to establish the multiphase SPH (Tartakovsky and Meakin 2005; Zhou et al. 2008; Breinlinger et al. 2013; Szewc et al. 2013; Robinson et al. 2014; Tong and Browne 2014; Wang et al. 2016; Koch et al. 2017; Ray et al. 2017; Li et al. 2018). In the multiphase SPH, two distinct phases are discretized into two sets of SPH particles with different properties and the interphase interaction is implicitly described by the different SPH parameters corresponding to the two phases or expressed by the pair-force. It is the most general approach to simulate a two-phase flow with SPH. Different from the Lagrangian solver for TFM, the multiphase SPH approach employs an implicit interaction to couple the two phases and considers the solid as a collection of SPH particles.

The multiphase SPH method is the natural extension of the SPH method for multiphase flow simulation. It contains features and advantages of the SPH method in single-phase flow simulations and introduces various two-phase interaction (surface tension) models to accomplish multiphase flow simulations with a fine and accurate representation of the interfacial behaviors.

#### 3.2.1 For liquid-liquid systems

Simulations of liquid–liquid flows were earlier and frequent applications of the multiphase SPH. For such systems, the properties of the two fluids, e.g., density and viscosity, are generally not much different.

The Rayleigh–Taylor (R-T) instability is an instability that develops and evolves at the interface between two horizon parallel liquids with different properties. Cummins and Rudman (Cummins and Rudman 1999) first reported a SPH simulation of the R-T instability with an artificial surface tension. Tartakovsky and Meakin (Tartakovsky and Meakin 2005) developed a SPH method based on the number density to eliminate this artificial surface tension between the two fluids. More realistic behaviors at the interface (Fig. 7) and of the R-T instability (Fig. 8) were achieved.

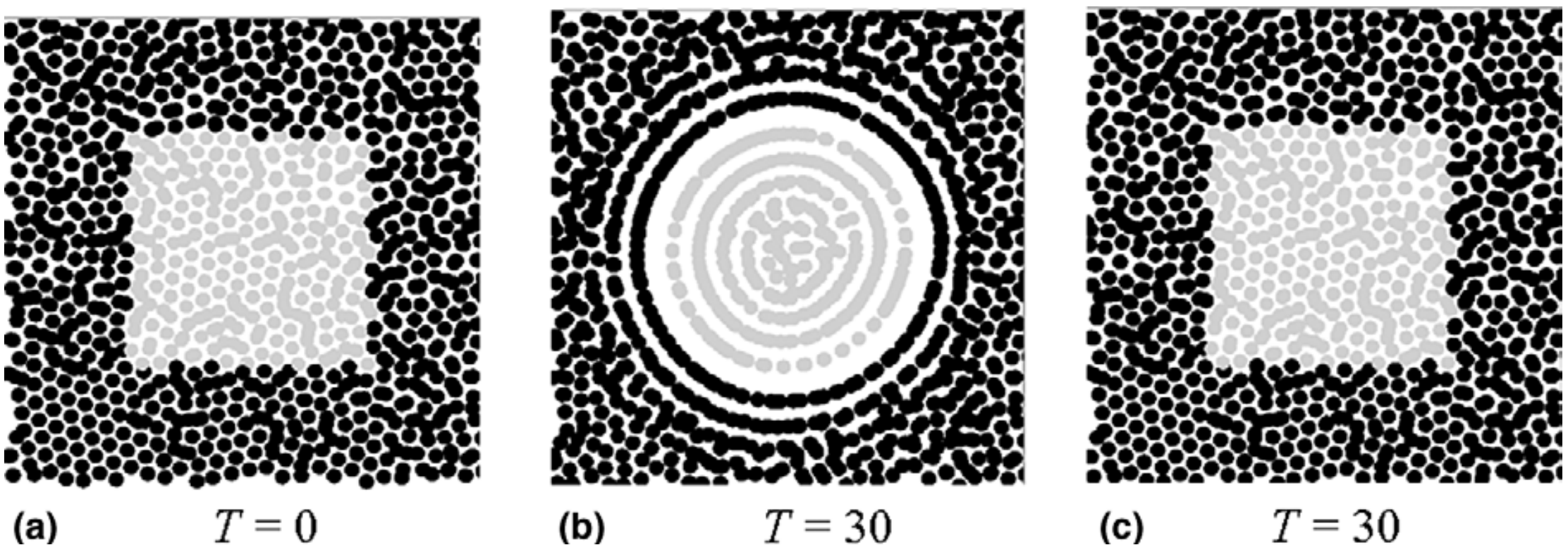

**Fig. 7.** Behavior of two fluids with different densities modeled with (b) standard and (c) modified SPH (Tartakovsky and Meakin 2005).

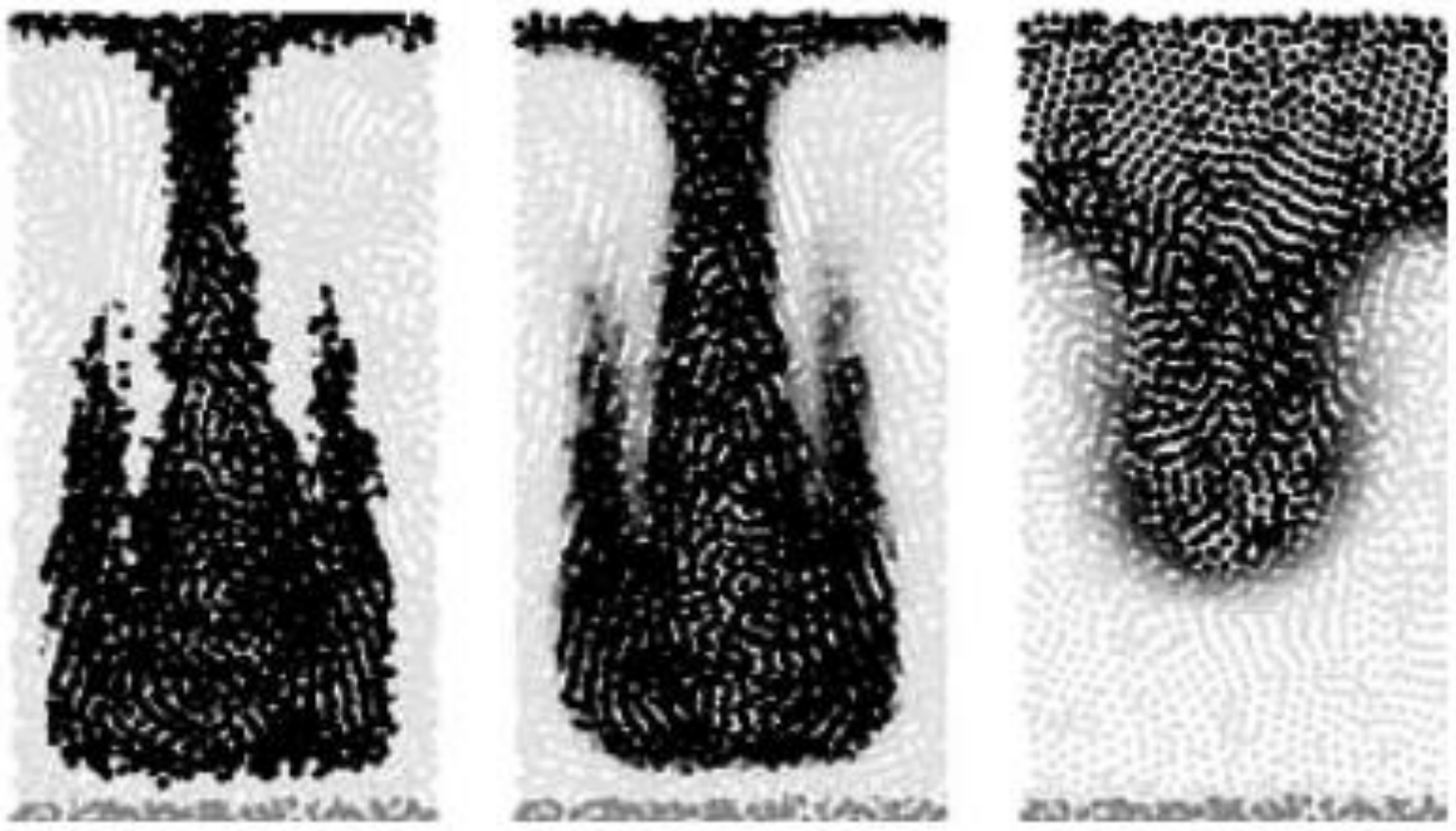

**Fig. 8.** Development of a two-dimensional Rayleigh–Taylor instability (Tartakovsky and Meakin 2005).

Hu and Adams (2007) developed the multiphase projection method to simulate the incompressible immiscible liquid–liquid flow. A SPH projection method with constant density was also developed and the R-T instability simulated using the CSF model (Hu and Adams 2009).

The Kelvin–Helmholtz (K-H) instability is another kind of instability of two horizontal parallel streams with different velocities. Using the multiphase SPH, Shadloo and Yildiz (2011) investigated the K-H instability (Fig. 9) of two immiscible incompressible fluids.

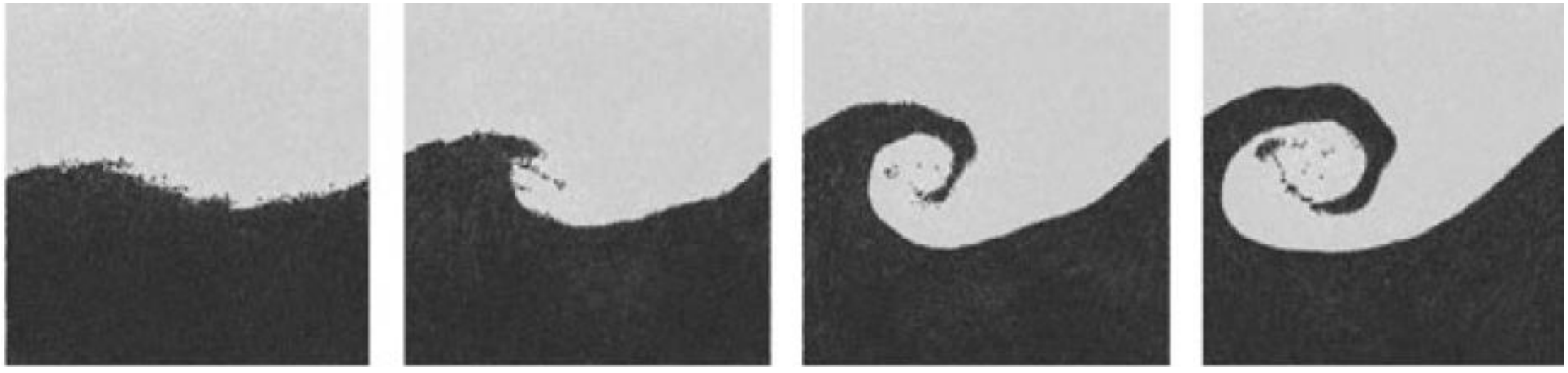

**Fig. 9.** Time evolution of the interface in a two-dimensional Kelvin–Helmholtz instability (Shadloo and Yildiz 2011).

Shao (2012) decoupled this two phase system using multiphase SPH and applied the pressure continuity condition at the interface. Tong and Browne (2014) used the SPH project method with constant density to investigate heat transfer between two fluids and the Marangoni force at the interface.

3.2.2 For liquid–gas systems

For a liquid–gas system, the density ratio can be much higher, of order hundreds or thousands. Dealing with this high-density ratio in the SPH method becomes a challenge in simulations. The standard SPH method cannot handle successfully implementations with density ratios higher than 10 (Colagrossi and Landrini 2003). Another difficulty with liquid–gas simulations using SPH is eliminating the pressure variation for stable surface tensions with high-density ratios.

Colagrossi and Landrini (2003) developed a corrective representation of the interface in multiphase SPH and simulated a dam breach in air. Das and Das (2009) developed a multiphase SPH with the CSF model for surface tension and the corrective kernel approximation to simulation bubble evolution from an orifice in a liquid pool.

Szewc (2012) developed the multiphase ISPH with density corrections and a color-function CSF for surface tension. A rising bubble in a liquid [Fig. 10(a)] with a density ratio of 1000 was simulated (Szewc et al. 2013). The $C_D$–$Re$ correlation was plotted, [Fig. 10(b)]. Lind et al. (2016) employed the multiphase ISPH to simulation the droplet deformation in gas with the density ratio 1000.

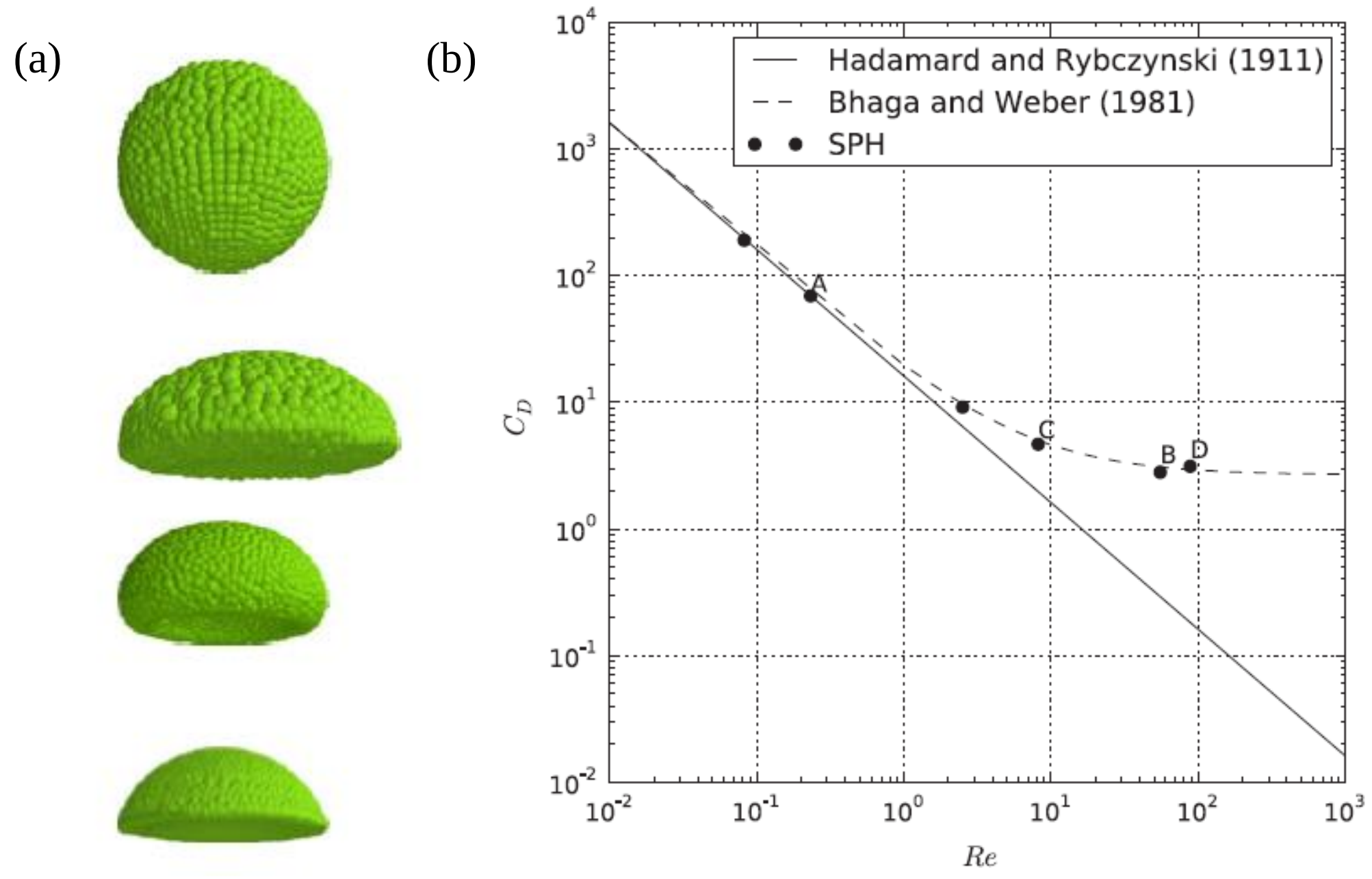

**Fig. 10.** (a) Bubble shapes rising in a liquid; (b) Relationship between coefficient of drag and Reynolds number (Szewc et al. 2013)

3.2.3. For liquid-solid systems

The solid in the multiphase SPH is mostly represented as a flow boundary or rigid body using SPH particles of quite different properties. Akinci et al. (2012) proposed a two-way coupling method for fluids and a rigid object using hydrodynamic forces in the multiphase SPH. Simulations of a boat passing a bridge and a frigate sailing in wavy seas were implemented.

Tofighi et al. (2015) represented solids as collections of SPH particles with very high viscosity to simulate the motion of rigid bodies in fluids. Fourtakas and Rogers (2016) proposed a model combining the yielding, shear, and suspension layers involved in the interaction of liquid and solid subject to adhesive effects. This multiphase SPH was employed to simulate breaching of an erodible dam.

3.2.4. For gas–solid systems

Ma et al. (2006) employed the macro-scale pseudo-particle method (MaPPM, a variant of SPH) to implement a high-resolution simulation of a gas–solid suspension. In the MaPPM, the solid particle is considered as a cluster of "frozen SPH particles". Xiong and collaborators (2010; 2011) used the MaPPM to perform a large-scale simulation of the gas–solid flow. Up to 30,240 solid particles and more than one billion fluid SPH particles were simulated with sub-grid precision (Fig. 11). This implementation was achieved on parallel GPUs.

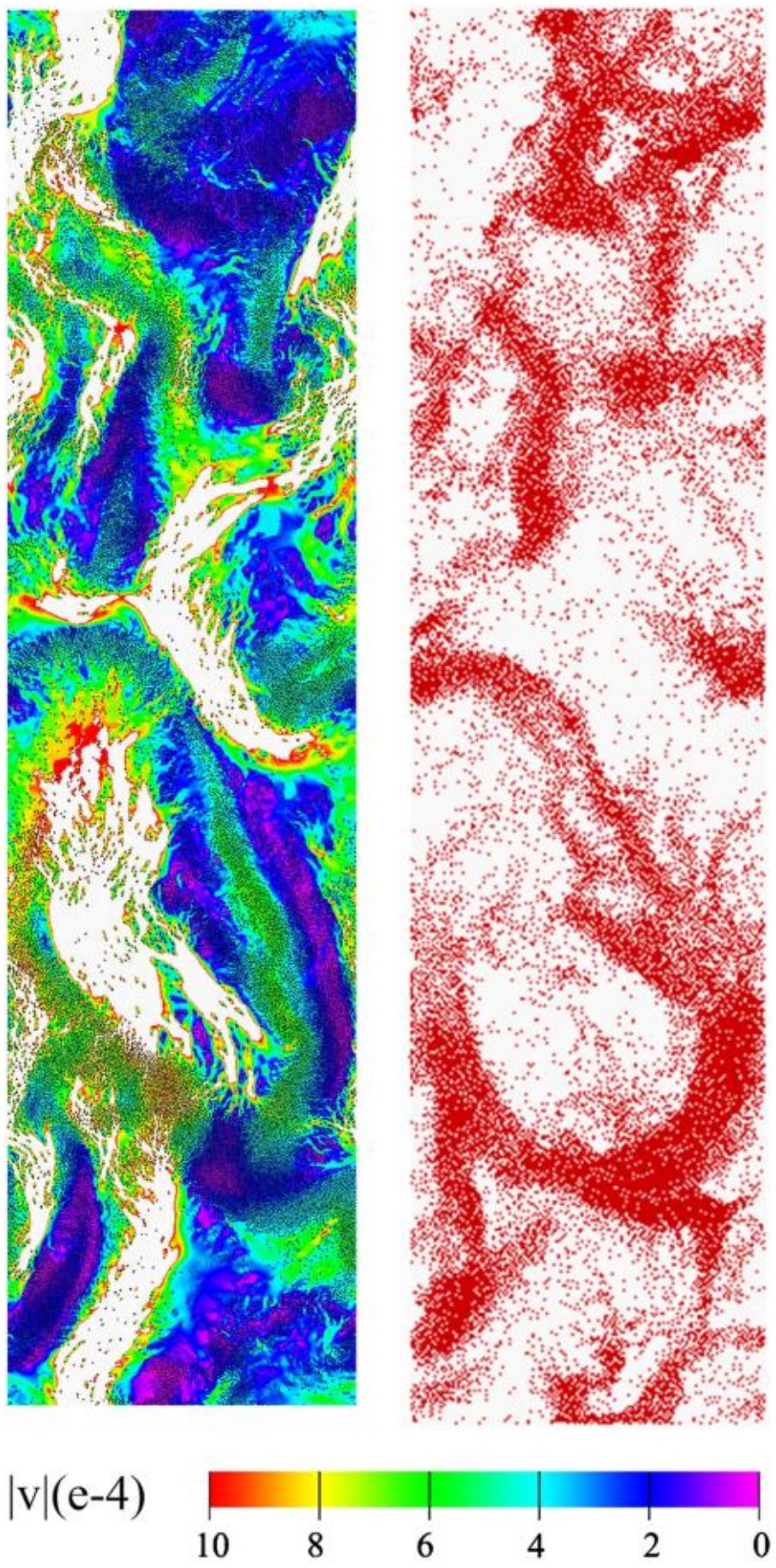

**Fig. 11.** Instantaneous flow field (left) and particle distribution (right) of a gas–solid system with 30,240 solid particles (Xiong et al. 2010).

### *3.3. Coupling of SPH and discrete methods*

Generally, in multiphase flows, SPH is suitable in solving a continuous phase or a phase with large deformation. The coupling of SPH and other discrete methods can be established easily to simulate fluid–solid systems and deformable bodies in a fluid. In coupling, SPH and other particle methods combine to benefit from their specific area of applicability. The coupling extends the applications of the SPH method to multiphase flows with different characteristics. Moreover, the advantages of the Lagrangian method can be incorporated in such simulations.

For fluid–particle systems, it is natural to trace the individual solid particles using the discrete element method (DEM). The interactions of the SPH particle and the DEM particles, i.e., drag and buoyancy forces, are formulated as functions of the difference in physical quantities between two phases in contact. Coupling of the SPH and the DEM achieves simulations of fluid–particle suspensions with an accuracy of a direct numerical simulation.

In other work, Potapov et al. (2001) presented a combined SPH-DEM simulation to investigate the flow around a cylinder and the two-dimensional fluid-solid flow between two parallel shearing plates. Huang and Nydal (2012) coupled the SPH and DEM to investigate the motion of solid particles in lid-driven cavity flow. Robinson et al. (2014) presented a SPH-DEM simulation of particle sedimentation in a 3D fluid column. In the simulation, the force on a solid particle by the fluid is express as

$$\boldsymbol{f}_i = V_p \left(-\nabla p + \nabla \cdot \tau\right)_i + \boldsymbol{f}_d \left(\theta_s, \boldsymbol{u}_f - \boldsymbol{u}_i\right), \tag{58}$$

where $V_\mathrm{p}$ is the volume of particle, and $\boldsymbol{f}_\mathrm{d}$ the drag force that acts on the particle and depends on the local solid volume fraction $\theta_\mathrm{s}$ and slip velocity between fluid and solid particle $\boldsymbol{u}_\mathrm{slip}=\boldsymbol{u}_\mathrm{f}-\boldsymbol{u}_i$. The pressure gradient and the stress tensor were evaluated using SPH interpolation,

$$\begin{aligned}
&\left(-\nabla p + \nabla \cdot \tau\right)_i = \frac{\sum_j m_j \theta_j W_{ij}\left(h_j\right)}{\sum_j m_j W_{ij}\left(h_j\right) / \rho_j} \\
&\theta_i = -\sum_j m_j \left[\left(\frac{p_i}{\Omega_i \rho_i^2} + \Pi_{ij}\right)\nabla_i W_{ij}\left(h_i\right) + \left(\frac{p_j}{\Omega_j \rho_j^2} + \Pi_{ij}\right)\nabla_i W_{ij}\left(h_j\right)\right]
\end{aligned} \tag{59}$$

The coupling force on the SPH particle is the weighted average of each DEM particle contribution,

$$\begin{aligned}
&\boldsymbol{f}_i = -\frac{m_i}{\rho_i}\sum_j \frac{1}{S_j} \boldsymbol{f}_j W_{ij}\left(h_c\right) \\
&S_i = \sum_j \frac{m_j}{\rho_j} W_{ij}\left(h_c\right)
\end{aligned} \tag{60}$$

Polwaththe-Gallage et al. (2016) employed the DEM-based spring network model to represent the viscoelastic membrane of red blood cells and used the SPH method to model fluids inside and outside of the cells. The coupled model was applied to simulate the deformation and motion of the cells in capillaries and to investigate influences associated with their interaction.

### 3.4. Coupling of SPH and grid-based methods

The SPH method is quite successful in fluid flow simulations, especially in interfacial flows, flows with large deformation, and multiphase flows. However, for high accuracy, much large numbers of computation particles are needed to represent the fluid, thus its computing demand is extremely high. For single-phase main flows, a full Lagrangian representation is not necessary and not efficient. For this reason, the coupling of the SPH and grid-based methods is employed to obtain a balance in accuracy and efficiency. In these coupling methods, the key is the treatment of the two approaches in their overlapping regions.

Marrone et al. (2016) developed a coupling method for SPH and finite volume (FV)-based CFD to simulate free surface flows with deformation and fragmentation. The main part is solved using structured cells and the upper part with large deformation is solved using SPH. Napoli et al. (2016) considered matching properties in the overlapping region of both FVM and SPH, and enhanced the accuracy of the simulation using the SPH-FV coupling method.

Deng et al. (2013) combined SPH and FV-based CFD to formulate the TFM for gas–solid flows. The dispersed solid particles were modeled as a fluid-like continuous phase and SPH was used to solve the solid phase using parallel GPUs; the fluid behaviors were solved using FV-based CFD implemented in the Ansys FLUENT software platform.

Feng et al. (2012) coupled SPH and the finite-element method to simulate the single acceleration process for abrasive water-jet cutting. Zhang et al. (2018) incorporated a large eddy simulation (LES) model in SPH to represent turbulent effects and used the finite-element fictitious boundary method to describe solid particles. The simulation of particle sedimentation in the fluid was performed to demonstrate the capability of the above improved SPH method in multiphase flow studies.

## 4. Miscellaneous discussions of SPH methods for multiphase flow

### *4.1. Corrections of approximation*

#### 4.1.1. Correction of kernel approximation

Chen et al.(Chen et al. 1999) proposed a corrective smoothed-particle method (CSPM) to address tensile instability by enhancing the accuracy in the kernel approximation,

$$W_{ij}^{CSPM}=\frac{\sum_j W_{ij}}{\sum_j \frac{m_j}{\rho_j} W_{ij}}. \quad (61)$$

CSPM was successfully used to improve the approximate solution at the boundary(Chen et al. 1999) and at the interface(Das and Das 2009; Tong and Browne 2014).

#### 4.1.2. Correction of particle approximation

Applying the mass density expression in the multiphase phase directly may induce an artificial surface tension and distort the density at the interface(Colagrossi and Landrini 2003). Tartakovsky and Meakin(Tartakovsky and Meakin 2005) proposed using the particle number density to eliminate the influence of different phase properties on the calculation of the density, thus avoiding the artificial surface tension. The particle number density $n$ is calculated using

$$n_i=\sum_j W_{ij}\ , \quad (62)$$

so that the momentum equation is then represented as

$$\frac{d\boldsymbol{u}_i}{dt} = -\frac{1}{m_i}\sum_j\left(\frac{p_i}{n_i^2}+\frac{p_j}{n_j^2}\right)\nabla_i W_{ij} + \frac{1}{m_i}\sum_j \frac{\mu_i+\mu_j}{n_i n_j}\boldsymbol{v}_{ij}\left(\frac{1}{r_{ij}}\frac{\partial W_{ij}}{\partial r_{ij}}\right)+\boldsymbol{g}\,. \tag{63}$$

Moreover, this approximation can be used to handle high-density-ratio multiphase flows.

4.1.3. Correction of consistency

The consistency ($C^n$ consistency) in SPH approximation is expressed as the order of a polynomial exactly reproduced by the SPH methods (Liu et al. 2005). It is influenced by the smoothing kernel function, smoothing length, boundary conditions and particle distribution etc. Generally, the conventional SPH methods can be considered restoring up to first order consistency ($C^1$ consistency). But for the cases with any feature of irregular particle distribution, smoothing length not same to particle spacing or boundaries (solid boundaries or interface between phases), even zero-th order ($C^0$) consistency cannot be exactly satisfied in conventional SPH methods (Liu and Liu 2006).

Liu and his co-workers (2005; 2006) proposed the finite particle method (FPM) to improve the consistencies of SPH methods. In FPM, the approximations of a field function $f(\boldsymbol{x})$ and its derivatives are corrected as (Liu et al. 2005)

$$\begin{Bmatrix} f_i \\ \nabla f_i \end{Bmatrix} = \boldsymbol{L}^{-1}\begin{bmatrix} \int f(\boldsymbol{x})W(\boldsymbol{x}-\boldsymbol{x}_i)d\boldsymbol{x} \\ \int f(\boldsymbol{x})W\nabla(\boldsymbol{x}-\boldsymbol{x}_i)d\boldsymbol{x} \end{bmatrix}, \tag{64}$$

where the corrective matrix $\boldsymbol{L}$ is written as

$$\boldsymbol{L} = \begin{bmatrix} \int W(\boldsymbol{x}-\boldsymbol{x}_i)d\boldsymbol{x} & \int(\boldsymbol{x}-\boldsymbol{x}_i)W(\boldsymbol{x}-\boldsymbol{x}_i)d\boldsymbol{x} \\ \int \nabla W(\boldsymbol{x}-\boldsymbol{x}_i)d\boldsymbol{x} & \int(\boldsymbol{x}-\boldsymbol{x}_i)\nabla W(\boldsymbol{x}-\boldsymbol{x}_i)d\boldsymbol{x} \end{bmatrix}. \tag{65}$$

The FPM greatly improves the consistencies of SPH methods for the cases encountered the problems of boundary deficiency and particle-distribution sensitivity. But for the cases with highly disordered particle distribution, the FPM simulation may be crashed due to the numerical instability leaded by the ill-conditioned corrective matrix (Zhang and Liu 2018). Reasonably assuming the contribution from self-direction is dominant for the derivatives of smoothing kernel function, Zhang et al. (2018) simplified the corrective matrix to a diagonal matrix as

$$\boldsymbol{L} = \begin{bmatrix} \int(\boldsymbol{x}-\boldsymbol{x}_i)Wd\boldsymbol{x} & 0 & 0 & 0 \\ 0 & \int(\boldsymbol{x}-\boldsymbol{x}_i)W_x^{'}d\boldsymbol{x} & 0 & 0 \\ 0 & 0 & \int(\boldsymbol{x}-\boldsymbol{x}_i)W_y^{'}d\boldsymbol{x} & 0 \\ 0 & 0 & 0 & \int(\boldsymbol{x}-\boldsymbol{x}_i)W_z^{'}d\boldsymbol{x} \end{bmatrix}, \tag{66}$$

to establish the decoupled finite particle method (DFPM). Thus, the particle approximation of $f(\boldsymbol{x})$ and its derivatives are written as

$$
\begin{aligned}
f_i &= \sum_j f_j W_{ij} \Delta V_j \Big/ \sum_j W_{ij} \Delta V_j \\
f'_{i,x} &= \sum_j \left( f_j - f_i \right) \partial W_{ij} / \partial x_i \, \Delta V_j \Big/ \sum_j x_{ji} W_{ij} \Delta V_j \\
f'_{i,y} &= \sum_j \left( f_j - f_i \right) \partial W_{ij} / \partial y_i \, \Delta V_j \Big/ \sum_j y_{ji} W_{ij} \Delta V_j \\
f'_{i,z} &= \sum_j \left( f_j - f_i \right) \partial W_{ij} / \partial z_i \Big/ \sum_j z_{ji} W_{ij} \Delta V_j
\end{aligned} . \qquad (67)
$$

*4.2. Tensile instability*

Tensile instability is a well-known problem in applying the SPH method, causing particle clustering or particle blowing away (Yang et al. 2014). It is frequently observed in SPH simulations and has been the focus of attention from early SPH studies (Schuessler and Schmitt 1981). Tensile instability still occurs even when the pressure is positive. Swegle et al. (1995) and Monaghan (2000) proposed a criterion for stability or instability for a stress state; the condition for unstable growth is given as

$$
W'' S > 0 , \qquad (68)
$$

where $W''$ is the second derivative of the kernel function, and $S$ the stress state. This condition shows that the kernel function is vital for stable simulations when using the SPH method. The second derivative of the kernel function should be non-negative in the entire support domain. Moreover, a hyperbolic-shaped kernel has been suggested to remove the tensile instability (Liu et al. 2003).

*4.3. Flows with high Reynolds number*

As is well known, similar to other particle methods, the conventional SPH method is not suitable to simulate flows under high Reynolds number, for tensile instability generated under high vorticity and non-real void generation under negative pressures (Sun et al. 2018).

Sun and colleagues (2017; 2018) implemented a tensile instability control (TIC) and particle shifting technique (PST) in the δ-SPH method (Antuono et al. 2012) to establish the $\delta^+$-SPH method. In the $\delta^+$-SPH method, an additional numerical diffusive term is introduced in the continuity equation and the particle shifting is applied to structure the particle distribution. Thus, the $\delta^+$-SPH method overcomes the limitations of the SPH simulations of flows with high Reynolds number.

*4.4. Surface tension coefficient*

As described in earlier sections, the surface tension at the interface between two phases is mainly represented using the CSF, IPF or CPM in SPH methods. In the CSF approach, the physical coefficient of surface tension is directly used to express the macroscopic surface tension force. For IPF and CPM approaches, however, the coefficient of surface tension is implicit in the model parameters and a theoretical relationship is yet to be attained (Tartakovsky and Meakin 2005).

Zhou et al. (2008) investigated the deformation of a drop in another fluid on a solid substrate with different parameter $s$ in the IPF model,

$$\boldsymbol{f}_{ij} = s\frac{\boldsymbol{r}_{ij}}{r_{ij}^{2}}. \tag{69}$$

The coefficient of surface tension $\gamma$ can be calculated using the sessile drop method according to

$$\gamma = \frac{1}{2}\Delta\rho g H^{2}, \tag{70}$$

where $\Delta\rho$ is the difference in density between the two liquids and $H$ the semi-height of the drop (Fig. 12). Thus, the relationship between the model coefficient $s$ and physical surface tension coefficient $\gamma$ is (Zhou et al. 2008)

$$\gamma = 1350.68 \times s^{0.501}. \tag{71}$$

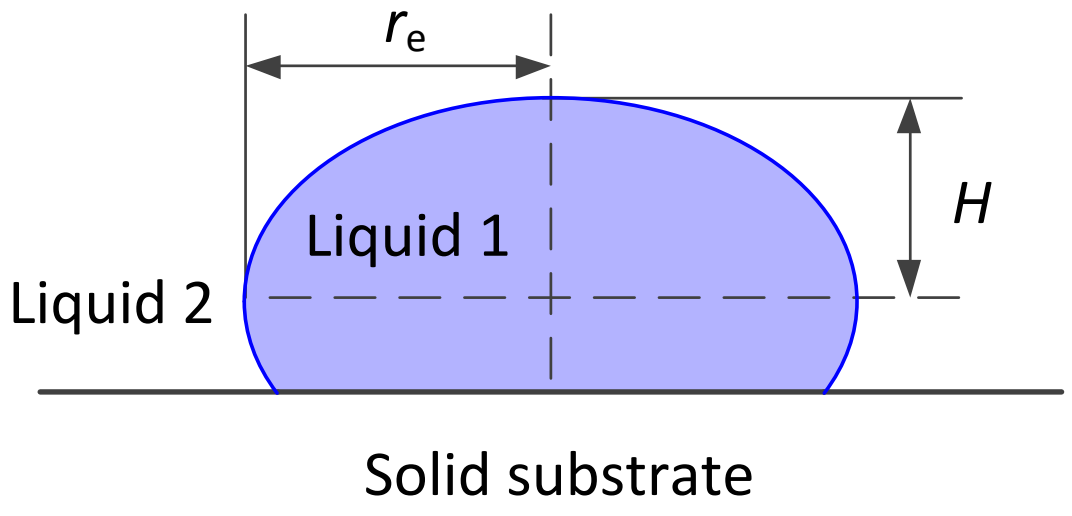

**Fig. 12.** Profile of a sessile drop (Zhou et al. 2008).

Using CPM, Xu et al. (2009) and Yang et al. (2014) applied the Laplace equation to calculate the surface tension from the SPH simulation for a viscous liquid drop with different radii,

$$p_l = p_g + \frac{\gamma}{R}, \tag{72}$$

where $p_l$ is the pressure at the center of the drop, $p_g$ the vapor pressure away from the drop, $\gamma$ the surface tension coefficient, and $R$ the radius of the drop. The particle pressure is computed from the particle density with the vdW EOS. The pressure at the center of the drop was linearly fitted to $1/R$. Hence, the coefficient of surface tension is identified as the gradient in the linear relationship; see Fig. 13.

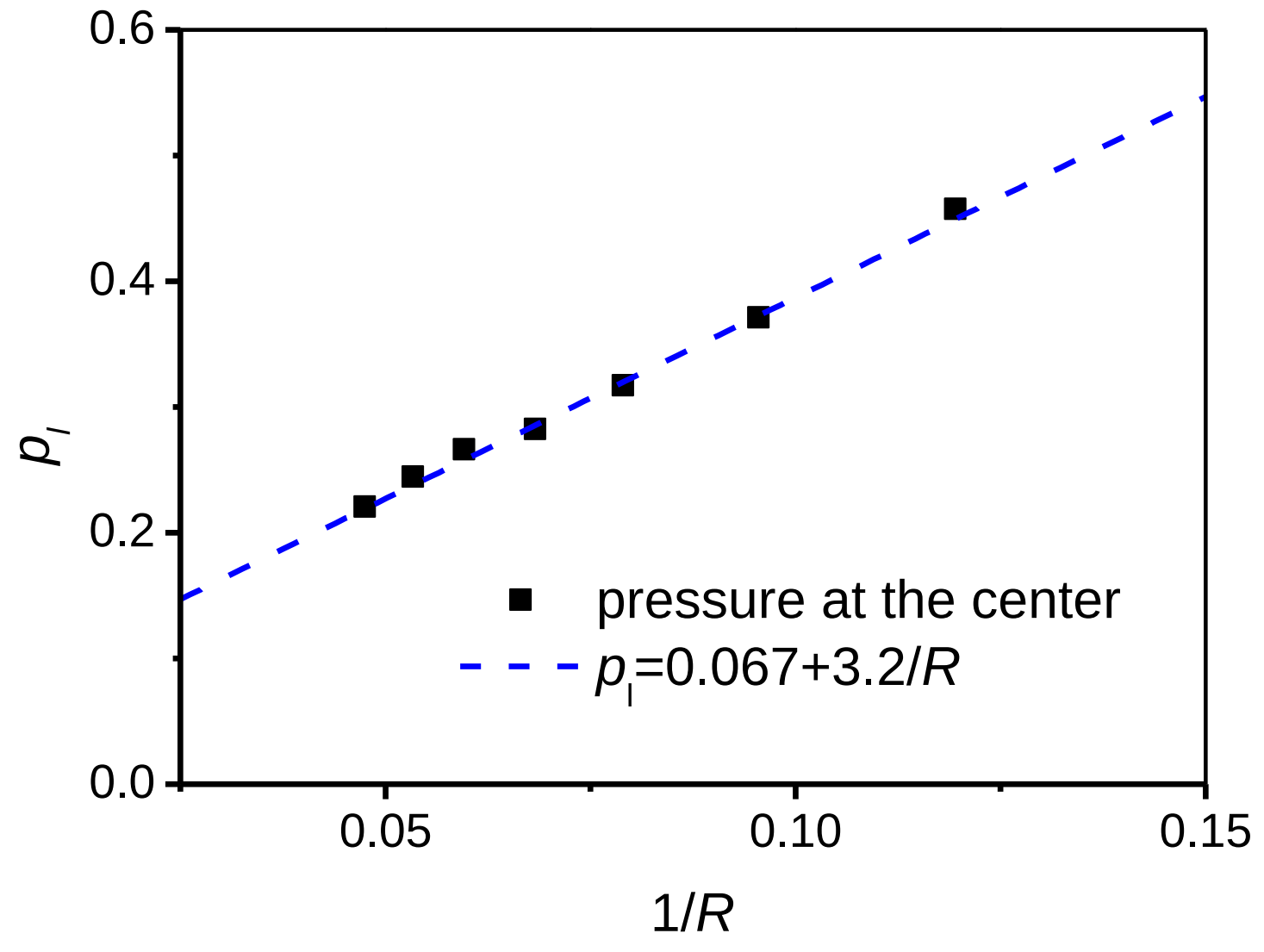

**Fig. 13.** Pressure at the center of drops for various drop radii (Yang et al. 2014).

## 5. Summary

As representative of meshfree methods, SPH has developed rapidly in the past decades and widely applied in many research areas. For its natural Lagrangian features, SPH, with its ability to deal with interfaces, large deformations, discontinuities, and multi-physics, has shown great potential in simulations of multiphase flows. Many researchers have contributed to improve multiphase flow simulations using the SPH method. This work attempted to summarize and classify these studies reporting SPH simulations of multiphase flow.

1) Instigated to solve the single-phase flow using SPH methods, the fully Lagrangian solver for the TFM is a straightforward way to simulate multiphase flows via the SPH method. However, explicit expressions arise for the coupling terms and modeling parameters (e.g., pressure and viscosity) of the dispersed phase in multiphase system.

2) The multiphase SPH method separately models each phase and naturally couples two phases with the different SPH parameters. The multiphase SPH method is the most applied of the SPH method for multiphase flow simulation and has been successfully applied to investigate a wide range of problems in disparate fields.

3) Also, the SPH method combined with other discrete methods has been employed to address multiphase flow simulations in coupled interaction, such as molecular dynamics, dissipative particle dynamics, and DEM. The coupling maintains the features and advantages of the discrete methods for each phase.

4) Furthermore, to reduce computation costs of multiphase flow simulations using large numbers of SPH particles, the simple main flow can be solved with grid-based methods. The Eulerian–Lagrangian coupling method provides a compromise between accuracy and efficiency for multiphase flow simulations and shows great potential for realistic multiphase flows of interest to industry.

However, to improve accuracy and efficiency of SPH simulations for multiphase flows, some problems remain to be addressed. For example, 1) Relationship between SPH parameters and physical properties.—The different weighted kernel function induces different effective properties (Monaghan 2006; Tartakovsky and Panchenko 2016). For steady simulations, several parameters such as artificial viscosity (Monaghan 1989) are introduced. Therefore, the calibration varies with different kernel functions, smoothing lengths, and approximations. 2) Solution to the high cost of computation.—While the SPH method is suitable to represent fluid behaviors with large deformation and distinct interfaces, its computation cost is critically high, especially simulations of multiphase flows. One solution is to implement large-scale parallel computing (Xiong et al. 2010; Fourtakas and Rogers 2016; Koch et al. 2017); another is to reduce the unnecessary computations by coupling with other discrete methods for dispersed bodies and grid-based methods for continuous phases in smooth flows. 3) Validation and augmentation from/to experiments.—There have been numerous experiments validating the SPH simulations, such as dam breaching (Koshizuka and Oka 1996; Ming et al. 2018) and gas–liquid atomization (Koch et al. 2017). In contrast, SPH simulations are widely applied in computer vision to display realistic fluid behaviors (Harada et al. 2007; Cornelis et al. 2014). The combination of experiment, simulation, and vision may incentivize enhanced virtual experiments on computers that are difficult or unachievable in reality.

## Acknowledgements

This work has been financially supported by the Science Challenge Project (No. TZ2016001), the Fund of State Key Laboratory of Multiphase Complex Systems (No. MPCS-2019-A-05), and the Strategic Priority Research Program of Chinese Academy of Sciences (No. XDA21030700). The author wishes to acknowledge the suggestions and comments on this paper by Prof. Moubin Liu (Peking University).